\documentclass[aps,preprint,amssymb,showpacs,groupedaddress]{revtex4-1}
\usepackage{graphicx}
\usepackage{xcolor}
\usepackage{amsmath}
\def\beq{\begin{equation}}
\def\eeq{\end{equation}}
\def\bea{\begin{eqnarray}}
\def\eea{\end{eqnarray}}

\begin{document}


\title{Micromechanics of intruder motion in wet granular medium}


\author{Rausan Jewel}
\affiliation{Department of Physics, Clark University, Worcester, MA 01610 } 
\author{Andreea Panaitescu}
\affiliation{Department of Physics, Clark University, Worcester, MA 01610 } 
\author{Arshad Kudrolli}
\email[]{akudrolli@clarku.edu}
\affiliation{Department of Physics, Clark University, Worcester, MA 01610 }

\date{\today}

\begin{abstract}
We investigate the effective friction encountered by an intruder moving through a sedimented medium which consists of transparent granular hydrogels immersed in water, and the resulting motion of the medium. We show that the effective friction $\mu_e$ on a spherical  intruder is captured by the inertial number $I$ given by the ratio of the time scale over which the intruder moves and the inertial time scale of the granular medium set by the overburden pressure. Further, $\mu_e$ is described by the function $\mu_e(I) = \mu_s + \alpha I^\beta$, where $\mu_s$ is the static friction, and $\alpha$ and $\beta$ are material dependent constants which are independent of intruder depth and size. By measuring the mean flow of the granular component around the intruder, we find significant slip between the intruder and the granular medium. The motion of the medium is strongly confined near the intruder compared with a viscous Newtonian fluid and is of the order of the intruder size.  The return flow of the medium occurs closer to the intruder as its depth is increased. Further, we study the reversible and irreversible displacement of the medium by not only following the medium as the intruder moves down but also while returning the intruder back up to its original depth. We find that the flow remains largely reversible in the quasi-static regime, as well as when $\mu_e$ increases rapidly over the range of $I$ probed.  
\end{abstract}


\maketitle


\section{Introduction}
The motion of objects through wet granular materials consisting of athermal solids sedimented in a fluid medium is encountered in a range of chemical and food processing industries, besides the muddy bottoms of ponds and rivers~\cite{gray09,balmforth14}. In the quasi-static limit, the drag experienced by objects of various shapes, and their interactions, have been investigated in granular media to study fundamental granular physics and biolocomotion~\cite{pacheco10,hosoi15,maladen10,reddy11,bergmann17,jslonaker17}. Further, drag experienced by an intruder moving in two and three dimensions well above the quasi-static regime has been also investigated in dry granular materials in gravity to find appropriate scaling laws~\cite{katsuragi07,katsuragi13,hilton13,takehara14,takada16,kumar17}.  However, the presence of the fluid changes the physics of the system considerably because it introduces drag, lubrication and pore pressure into the system~\cite{happel,stevens05,delannay17}.  Viewed from the perspective of fluids, the presence of athermal frictional grains in the medium makes the physics of the problem also completely different from that of an intruder moving in a viscous fluid~\cite{lauga09}.

Intruder dynamics in wet granular medium is doubly challenging because the rheology of the medium is not well understood, and {\color{black} the flow around the intruder is time-independent, i.e.} unsteady. The intruder causes transient fluidization of the athermal medium which is otherwise static. The sedimented granular medium considered here are theoretically distinct from granular suspensions where the grains are also athermal and can come into frictional contact~\cite{seto13,brown14}, but where grains have the same density as the fluid and can be considered to be uniformly distributed unless shear gradients are present. Moreover, the momentum exchange between the fluid and the granular phase in the medium is also different when the medium is sheared because of the density difference. Thus, the presence of solids leads to considerable differences from the motion of a particle sedimenting through a Newtonian fluid, or for that matter when particulates are present in small concentrations~\cite{happel}. 

Recently, it was demonstrated~\cite{apanaitescu17} that a sphere dragged through granular hydrogels immersed in water can be described by an effective friction which scales with inertial number $I$~\cite{cruz05}, and increases non-linearly from a non-zero static value. The form was found to be similar to that derived from the Herschel and Bulkley model~\cite{herschel26}, which is used to describe non-Newtonian fluids and muds~\cite{hemphill93}. 
Building on that study, we probe the dynamics of an intruder settling through granular hydrogels immersed in water as a model of wet granular medium or mud consisting of soft granular medium immersed in water. This is a much simplified system compared to experiments on intruders settling in clay and cornstarch suspensions which are more difficult to probe experimentally, as they show further complex material dependence as well~\cite{gray09,gueslin09,kann11}.  

Exploiting the near transparency of the granular hydrogel medium, we visualize the motion of the intruder as it accelerates, after being released from rest, and extract the encountered effective friction. To understand the relation between the observed rheology and the micromechanics of the medium, we visualize the motion of the medium around the intruder by adding tracer particles. We show that the flow of the medium is strongly confined around the intruder, and different than that for a viscous fluid. We then describe the effect of intruder speed and depth on the rearrangement of the medium, and its reversibility as a function of inertial number. 
    

\section{Experimental System}
\begin{figure}
\begin{center}
\includegraphics[width=.65\textwidth]{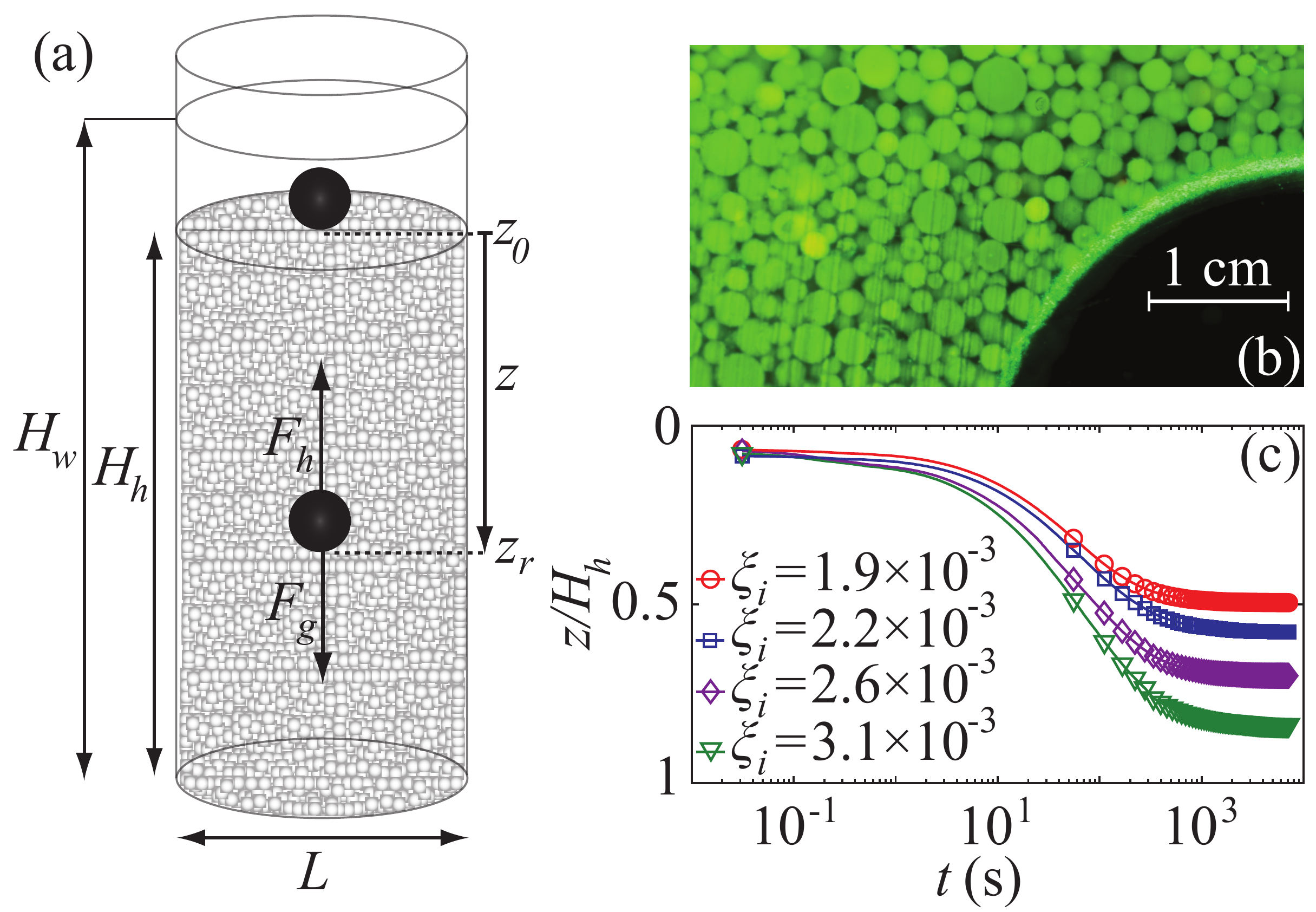}
\end{center}
\caption{(a) Schematic of the experimental system consisting of an intruder descending through sedimented granular hydrogels immersed in water. The depth $z$ of the intruder is measured from the top of the sedimented hydrogels denoted by $z_o$ to the bottom of the intruder. The height of the water column $H_w$ and the  granular hydrogel medium $H_h$ in the container are also shown. (b) A transect of the sedimented medium and the intruder illuminated by a thin laser sheet. (c) The intruder depth $z$ as a function of time $t$ for a range of  $\xi_i$, intruder density relative to the medium. 
}
\label{fig:experimental}
\end{figure} 

Fig.~\ref{fig:experimental}(a) shows the system used to investigate the settling dynamics of the spherical intruder in a container filled with a granular medium sedimented in a fluid. The grains are composed of hydrated polyacrylamide with diameters $d_h = 1.5 \pm 0.5$\,mm,  density $\rho_h = 1004\,$ kg\,m$^{-3}$, Young's modulus $E \sim 10$\,kPa, friction coefficient $\mu_h \sim 10^{-2}$  similar to previous work~\cite{mukhopadhyay11,apanaitescu17}. The grains sediment to the bottom of the container to a height $H_h$ filled with distilled water. Water is filled to a height $H_w > H_h$ in all our experiments to prevent surface tension effects from playing any role in the observed phenomena. Typically,  we use a cylindrical container with $H_w = 430$\,mm and $H_h = 380$\,mm, and horizontal width $L = 180$\,mm. These dimensions are chosen to be sufficiently large to be unimportant to the dynamics studied. 

The grains are visualized by using a thin illumination sheet generated by a laser and cylindrical lens combination, and appear to be randomly packed (see Fig.~\ref{fig:experimental}(b)).  By measuring the volume of water displaced, the volume fraction of the grains in the medium is found to be $0.6$. Both the random packing and volume fraction are consistent with typical spherical grain packings obtained at high deposition rates~\cite{apanaitescu14}, and lower than packing fractions found with frictionless spheres~\cite{silbert02}. The density of the hydrogel medium $\rho_m$ {\color{black} and the density of water $\rho_w$  is found to be $1001 \pm 1$\,kg\,m$^{-3}$ and $998 \pm 1$\,kg\,m$^{-3}$ respectively} at 24$^o$C.   The change in their volume due to the overburden pressure 
\begin{equation}
{\color{black}
P_p = (\rho_h - \rho_w) g z,
}
\label{eq:pressure}
\end{equation} 
where $z$ is the depth measured from the bed surface  \textcolor{black} {$z_0$ to the depth $z_r$  where, the intruder comes to rest}, can be estimated assuming linear elasticity to be less than 0.01\% at the deepest point $z = H_h$ in the container. We thus assume that the density of the hydrogel medium 
\begin{equation}
{\color{black}\rho_m = \phi\rho_h+\left(1-\phi\right)\rho_w}
\end{equation}
is essentially constant throughout the system for the purpose of our study. 

The intruders used in our studies consist of spherical shells, with diameter $d_i = 27$\,mm, 40\,mm, and $50$\,mm filled with various amounts of glass beads to vary their density $\rho_i$ without changing their size and surface properties. The relative density difference between the intruder and the hydrogel medium is then given by $\xi_i = \rho_i/\rho_m -1$, {\color{black}where $\rho_i$ is the density of the intruder} and the values of $\xi_i$ are listed in Tab.~\ref{tab:dens} corresponding to the various intruders. Because the hydrogels are essentially transparent and have a refractive index close to that of water, we can visualize the position of the intruder inside the medium using back lighting and a digital camera. A movie of an intruder as it falls through the medium can be found in the supplementary documentation~\cite{supdoc}. 
The intruder is located by identifying the centroid of the dark pixels associated with the intruder to within $\pm 0.5$\,mm or less than $\pm 0.01 d_i$ in the case of $d_i =5$\,cm. Then, the depth of the intruder $z$ is recorded from the surface of the sedimented hydrogel bed down to the bottom of the intruder. We use a well defined protocol to initialize the medium to obtain consistent results by stirring the granular hydrogel medium for a minute and allowing them to settle for 20 minutes before performing measurements to avoid the initial transients. 

\begin{table}
\begin{center}
\begin{tabular}{c c c}
\hline
$d_i$ (cm) & $d_i/d_h$ & $\xi_i$ ($\times 10^{-3}$)\\
\hline
2.7 & 18.3 &\,\,\,\,\,\,\,\, 0.6, 1.8, 2.2, 3.8, 5.8, 7.1  \\
4.0 & 26.7 &\,\,\,\,\,\,\,\, 2.4, 2.9, 3.1, 3.4, 3.6, 3.9 \\
5.0 & 33.3 &\,\,\,\,\,\,\,\, 1.1, 1.9, 2.02, 2.2, 2.6, 3.1 \\
\hline
\end{tabular}
\end{center}
\caption{List of intruder sizes and their density difference relative to the medium used in the measurements.}
\label{tab:dens}
\end{table} 

\section{Intruder Probed Rheology}
Fig.~\ref{fig:experimental}(c) shows the measured depth $z$ of intruders with various $\xi_i$  as they descend individually through the granular medium after being released from rest at the surface of the medium at time $t=0$\,s.  The data here is scaled with respect to the medium height $H_h$ to give a sense of the location of the intruder with respect to the container bottom which then corresponds to $z/H_h =1$.  In all cases, the intruder is observed to descend rapidly at first before slowing down, and then creeping for hours, before finally coming to rest. One can note that the intruder comes to rest at depth $z_r$ well above the container bottom as the intruder density is increased over the range of $\xi_i$ shown. (We monitored the intruder also over days in a few cases and found that the intruder fluctuates in place to within a fraction of the grain size which we attribute to small variation in the room temperature which can cause expansion and contraction to the grains and the container.) The intruder reaches the bottom of the container at a $\xi_i$ higher than Tab.~\ref{tab:dens}. Because the density of the medium is essentially constant with depth and $\xi_i > 0$, we infer that the intruder is held in place, because the medium exhibits a yield stress which needs to be exceeded for the intruder to move.

\subsection{Statics}

\begin{figure}
\begin{center}
\includegraphics[width=0.65\textwidth]{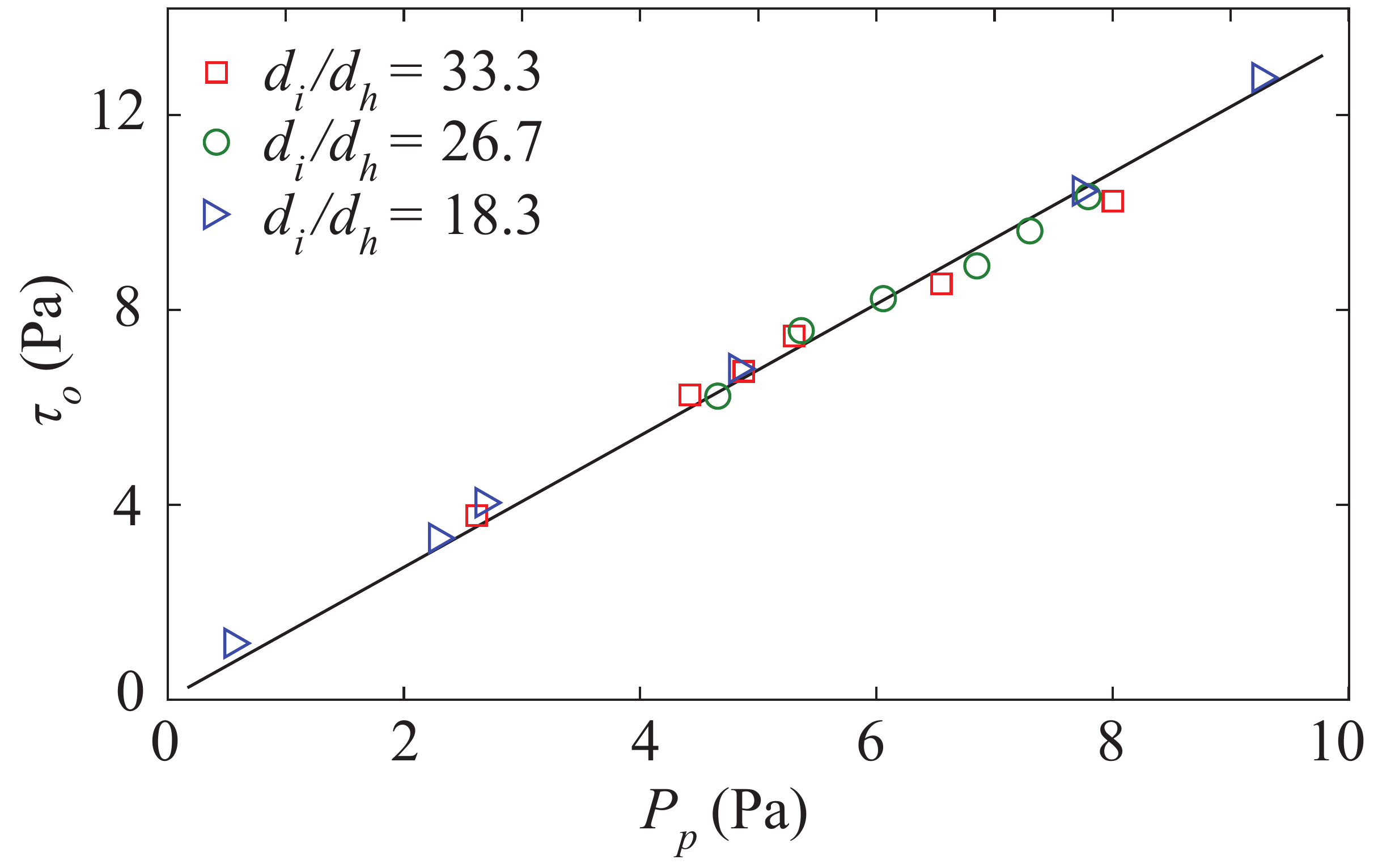}
\end{center}
\caption{The yield stress $\tau_o$ as a function of overburden pressure $P_p$ for various combinations of intruder densities and size. The slope corresponds to the effective static friction $\mu_s$. The error bars are the same as the symbol size and thus not drawn separately. }
\label{fig:statics}
\end{figure} 

We obtain the depth at which the intruder comes to rest $z_r$ as a function of relative excess density of the intruder $\xi_i$, and then estimate the stress applied by the intruder $\tau_o \sim F_g/A_i$, where $F_g$ is the force due to the non-buoyant weight of the intruder given by $F_g = \pi (\rho_i - \rho_m) g d_i^3/6$, and $A_i$ is the area over which $F_g$ is distributed. Because $d_i \gg d_h$, we assume that $A \sim \pi d_i^2/4$, and therefore 
\begin{equation}
\tau_o \approx \frac{2}{3}(\rho_i - \rho_m) g d_i. 
\end{equation}
Further, the overburden pressure $P_p$ due to the weight of the hydrogels at the depth where the intruder comes to rest is given by Eq.~\ref{eq:pressure}. 

Fig.~\ref{fig:statics} shows $\tau_o$ plotted versus $P_p$ corresponding to various intruder density and size. We observe that $\tau_o$ grows linearly with $P_p$, and all the data collapse onto a single line. {\color{black} Accordingly, one can define a coefficient of static friction $\mu_s$ corresponding to the ratio of the stress acting on the intruder in the direction of motion and the normal stress in the perpendicular direction similar to Ref.~\cite{apanaitescu17}. At the point where intruder has just come to rest, we assume that the stress acting on the intruder then just equals the yield stress of the medium. Because the intruder was moving in the downward direction we assume this stress is in the vertical direction. Then, in considering the normal stress, we make the assumption that the overburden pressure $P_p$ due to the weight of the grains above is approximately isotropic. Thus, we assume that the normal stress acting on the intruder in the horizontal direction is thus $P_p$ as well.  
}
Thus, the slope of the plot shown in  Fig.~\ref{fig:statics} corresponds to the $\mu_s$, {\color{black} given by
\begin{equation}
\mu_s = \frac{\tau_o}{P_p}\,,
\label{eq:mus}
\end{equation}
similar to the definition proposed in Ref.~\cite{apanaitescu17}. However,} care should be exercised when interpreting this definition in terms of internal friction angles of the medium because of the differences in prefactors associated with the geometry of the intruder. Here, we simply use this definition to characterize and nondimensionalize the drag experienced by the intruder with respect to the other force important in the problem. From the fit, we find $\mu_s = 1.3 \pm 0.02$. Hence, the observed $\mu_s$ are constant within experimental errors due to the residual variation in the room temperature rather than intruder depth measurement errors.    

It is noteworthy that the linear dependence of the yield stress with depth observed in Fig.~\ref{fig:statics}(a) is consistent with the study of Brzinski III {\it et al.}~\cite{Brzinski13} performed with an intruder penetrating a dry granular bed. There, it was shown that granular materials exert a force on the intruder which is locally normal to the surface of the object, while the tangential contributions are much smaller. In addition, normal forces increase with the gravitational loading pressure of the medium. With these two assumptions, the total force acting on a spherical intruder immersed in a dry granular medium was found to increase linearly with the depth of the intruder. Thus, our experiments reveal that in the static limit, the wet granular medium composed of granular hydrogels sedimented in water behave similar to dry granular medium with frictional contacts.


\subsection{Dynamics}

\begin{figure}
\begin{center}
\includegraphics[width=0.65\textwidth]{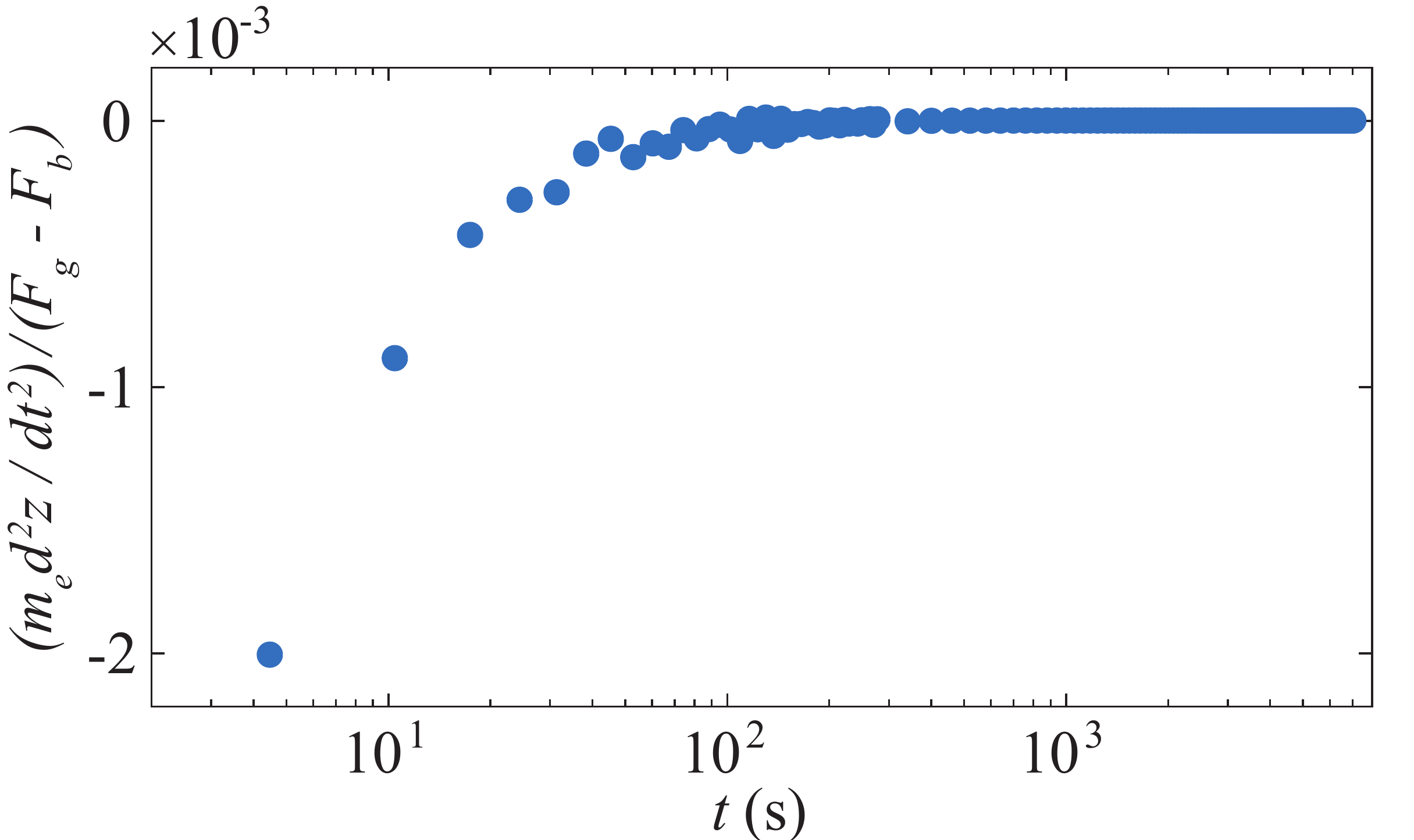}
\end{center}
\caption{The ratio of the relative magnitude of the inertial term, and the gravitational terms becomes steadily small as the intruder slows down. The data corresponds to depth versus time curve shown for $\xi_i = 3.1 \times 10^{-3}$ in Fig.~\ref{fig:experimental}(c).}
\label{fig:inertia}
\end{figure}

We next focus on the friction experienced by the intruder during the dynamic settling phase. Considering the mean forces acting on the intruder, we have 
\begin{equation}
F_d = F_g - F_b -  m_{e}\frac{{\rm d}^2z}{{\rm d}t^2},
\label{eq:force}
\end{equation}
where, $F_d$ is the drag force acting on the intruder, $F_g = \frac{\rho_i \pi d_i^3 g}{6}$ the gravitational force and $F_b = \frac{\rho_h \pi d_i^3g}{6}$ is the buoyant force due to the medium displaced, $m_{e}$ its effective mass which depends on the density of the intruder and the medium, and $\frac{{\rm d}^2z}{{\rm d}t^2}$ is the acceleration of the intruder. 

Fig.~\ref{fig:inertia} shows an example of the ratio of the acceleration term in Eq.~\ref{eq:force} divided by $(F_g - F_b)$ to understand the relative strength. Here, the added mass effect has to be included in any estimate of the effective mass of the intruder $m_e$ because $\rho_h \approx \rho_w$. Thus, $m_e \textcolor{black} {\approx} (\rho_i + \rho_m/2)\pi/12d_i^3$, where we have used a form of added mass correction in a Newtonian fluid. With this assumption, the effective mass can be estimated to be approximately 1.5 times the mass of the intruder. Except at very early times, when the intruder begins to accelerate from rest near the medium surface and overburden pressure is small, the relative strength is relatively very small. Thus, the acceleration term is small as the  intruder slowly comes to rest. Nonetheless, we include this correction in general in estimating $F_d$. {\color{black} Now, the drag force encountered by the moving intruder is proportional to the shear stress, due to the effective friction acting on a local surface element of the intruder, integrated over its entire surface area. However, for simplicity, we approximate the effective shear stress as the drag force divided by the cross section
of the intruder. Therefore,} we divide $F_d$ by the cross section area of the intruder $A_i$, and the overburden pressure as in the static case, to now obtain the effective friction $\mu_e$ as a function of the velocity $v_i$ of the intruder as it descends, through the medium, i.e. 
\begin{equation}
\mu_e = \frac{F_d/A_i}{P_p}\,, 
\label{eq:mue}
\end{equation}
where, we have made the same assumption as in obtaining the static effective friction given by Eq.~\ref{eq:mus} that the stress exerted on the intruder, in the directional normal to its motion, is approximately given by the overburden pressure $P_p$.

\begin{figure}
\begin{center}
\includegraphics[width=0.45\textwidth]{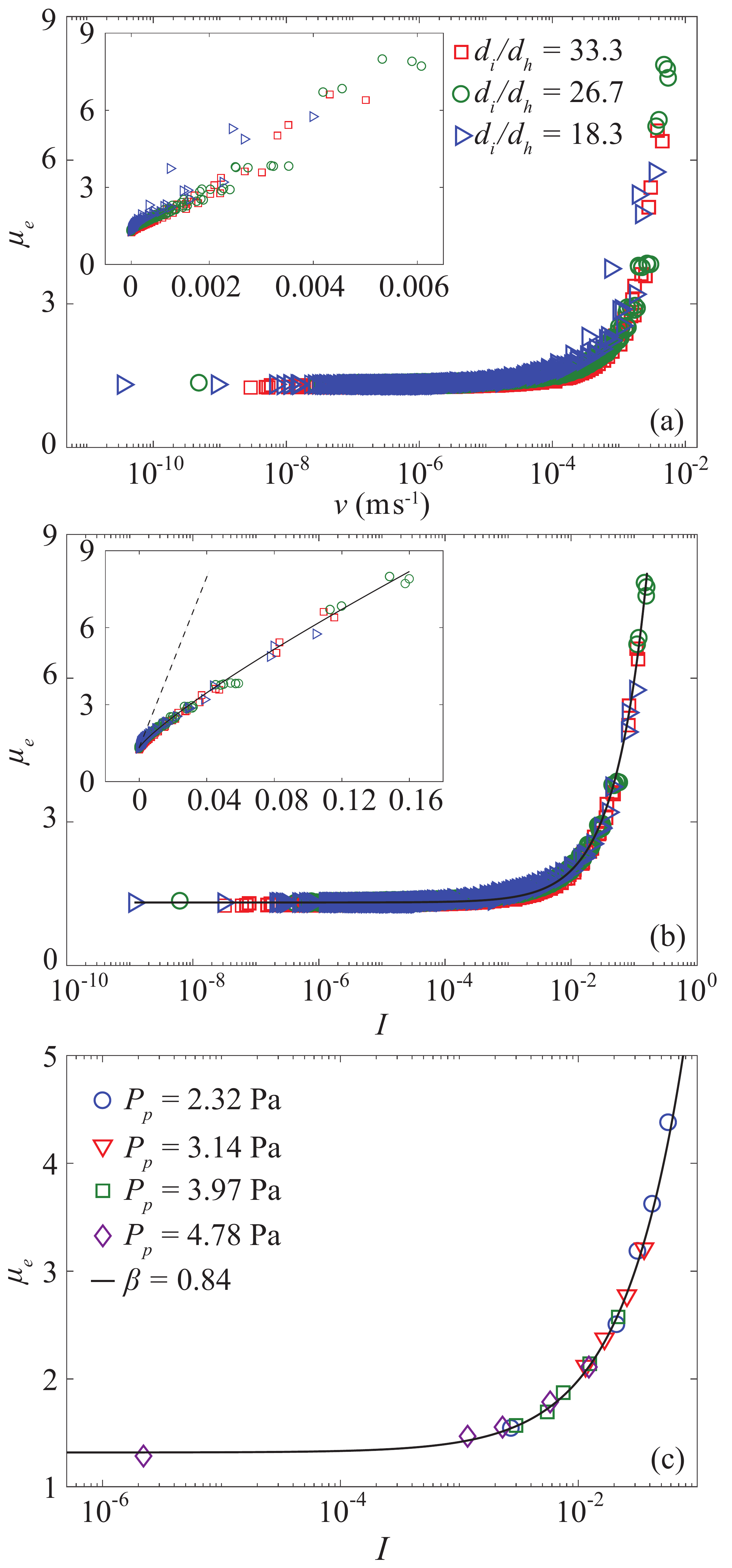}
\end{center}
\caption{(a) The effective friction $\mu_e$ as a function of intruder speed $v_i$ is observed to approach a constant value at low speeds. Inset: Same plot in linear scale. (b)  The effective friction $\mu_e$ as a function of inertial number $I$ along with Eq.~\ref{eq:muI}. Inset: Same plot in linear scale shows that the data collapses onto a single curve both at low and high velocities as a function of $I$. The key is the same as in (a). (c) $\mu_e$ as a function of inertial number $I$ for $d_i =5$\,cm for various depths. $\mu_e$ is observed to collapse onto the same curve, irrespective of depth. The measurement errors are smaller than the marker size and not drawn for clarity. }
\label{fig:dynamics}
\end{figure}

Fig.\,\ref{fig:dynamics}(a) shows $\mu_e$ probed by the intruder 
as a function of $v_i$ for various intruder sizes and relative densities listed in Tab.~\ref{tab:dens}. We observe from the log-linear form of the plot in Fig.\,\ref{fig:dynamics}(a) that the data at low velocities approaches a constant value. This is consistent with the findings in Fig.~\ref{fig:statics} that $\mu_s$ is observed to be constant, irrespective of the density and the size of the intruder. At higher speeds, we observe that $\mu_e$ increases in all cases but does not collapse onto a single curve. 

In Ref.~\cite{apanaitescu17}, it was shown that the drag experienced by an intruder as it moves with a constant speed $v_i$ is given by an effective friction which is only a function of the inertial number $I$, where $I$ is given by the time scale over which the intruder moves through its diameter and the inertial time scale set by the overburden pressure. Assuming that the shear rate of the medium can be estimated using the velocity of the intruder and its diameter, i.e. $v_i/d_i$, it was found that 
\begin{equation}
I = \frac{v_i}{\sqrt{P_p/\rho_h}}\,. 
\label{eq:I}
\end{equation}
Given that the original form of $I$~\cite{gmidi04} was defined using uniform shear conditions and constant shear rates, this interpretation and generalization to the unsteady flow conditions in the case of intruder dynamics is not a priori obvious. 

We plot the effective friction $\mu_e$ as a function of $I$ in log-linear and linear-linear format in  Fig.~\ref{fig:dynamics}(b). We observe that the data collapses onto a single curve. Thus, we fit the functional form found in Ref.~\cite{apanaitescu17}
\begin{equation}
\mu_e(I) = \mu_s + \alpha\, I^\beta\,,
\label{eq:muI}
\end{equation}
where, $\alpha$ and $\beta$ are empirical constants. \textcolor{black}{The value of $\beta$  in particular can provide insight into the nature of the medium as probed by the intruder. This is a similar form to the Hershel-Bulkley model~\cite{herschel26} for stress and strain rate scaling since $\mu_e$ is proportional to the stress at a given depth, and $I$ is proportional to the shear rate. In that model, $\beta$ is called the consistency index with $\beta < 1$ corresponding to a shear-thinning fluid, and $\beta > 1$ corresponding to a shear thickening fluid. In the case where $\beta = 1$, the Hershel-Bulkley model reduces to the Bingham plastic model of a viscoplastic material, in which the medium behaves like a viscous fluid above yield with viscosity proportional to $\alpha$.} We observe that the data collapses onto the curve with the effective friction $\mu_e$ approaching a constant value $\mu_s = 1.3 \pm 0.02$ independent of the intruder size. Further, the fit to Eq.~\ref{eq:muI} yields $\alpha =32 \pm 1$, and $\beta = 0.84 \pm 0.01$~\cite{hemphill93}. \textcolor{black}{In this case, the value of $\beta$ suggests that the medium is shear-thinning}. Thus, the increase of friction with $I$ is sub-linear, as was also found in the previous experiments with an intruder dragged with constant speed in similar sized granular medium~\cite{apanaitescu17}.

\begin{figure}
\begin{center}
\includegraphics[width=0.65\textwidth]{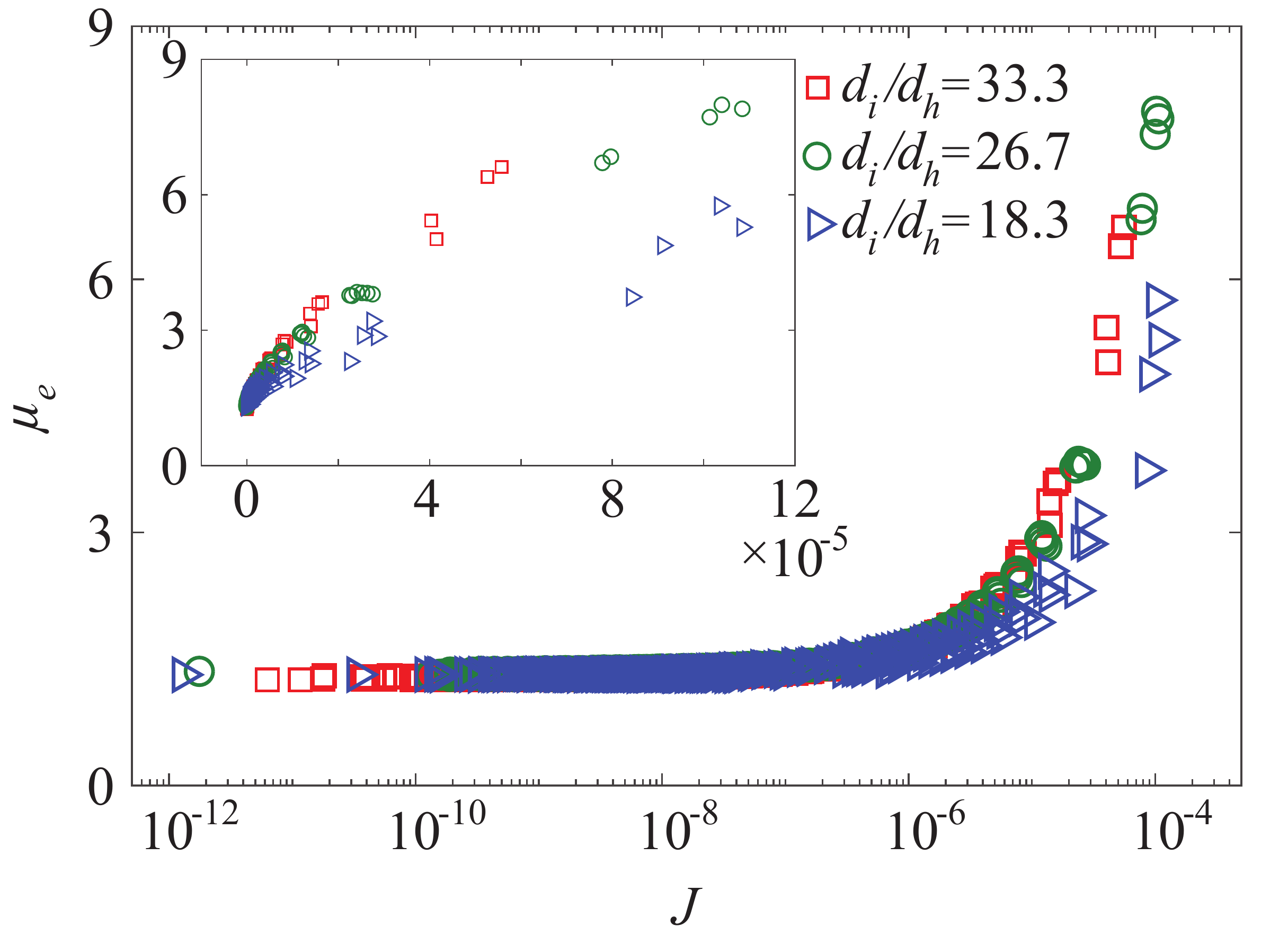}
\end{center}
\caption{Effective friction as a function of the viscous number $J$ is not observed to collapse onto a single curve. The observed scatter is far greater than measurement error which is smaller than symbol size.}
\label{fig:J}
\end{figure}

It is also noteworthy here that the observed $\mu_e(I)$ collapses onto the form with same $\alpha$ and $\beta$, irrespective of the depth of the intruder. To show this explicitly, we have plotted $\mu_e$ versus $I$ obtained at various depths, and thus $P_p$, in Fig.~\ref{fig:dynamics}(c). We observe that the data for all $z$ collapse onto same curve given by $\alpha$ and $\beta$ obtained to describe Fig.~\ref{fig:dynamics}(b).  

If one starts from the  Hershel-Bulkley relation given by $\tau = \tau_o + k \dot{\gamma}^\beta$~\cite{herschel26}, where $\tau$ is the shear stress, $\dot{\gamma}$ is the strain rate, and $k$ and $\beta$ are medium dependent constants, then, dividing by the overburden pressure $P_p$, and further assuming $\dot{\gamma} = v_i/d_i$, and rearranging in terms of $I$ using Eq.~\ref{eq:I}, we have $k = \alpha P_p^{1-\beta/2} d_i^\beta \rho_h^{\beta/2}$. Thus, given $\alpha$ is found to be independent of $P_p$, our measurements appear to imply a particular dependence of $k$ with depth in the case of the granular hydrogel medium.

In order to check if the observed evolution of the friction coefficient $\mu_e$ is determined by the viscous properties of the grains in the fluid, we examined a dimensionless viscous number $J$ in analogy with the one proposed for uniformly sheared neutrally buoyant suspensions~\cite{boyer11}, but by assuming a shear rate $\dot{\gamma}$ given by the speed of the intruder and its diameter just as in defining $I$ for our system. Thus, 
\begin{equation}
J =  \frac{\eta_s v_i}{P_p}\,,
\end{equation}
where, $\eta_s$ is the viscosity of the fluid, which in the case of our system is assumed to be $8.90 \times 10^{-4}$\,Pa\,s. 
We plot $\mu_e$ versus $J$ in log-linear style Fig.~\ref{fig:J}, and in linear-linear style in the inset. At low $J$, the measured $\mu_e$ is observed to converge to a constant value $\mu_s = 1.3$ as is also observed in Fig.~\ref{fig:statics}(b). However, the data does not collapse onto a single curve as it does in case of $I$ as shown in Fig.~\ref{fig:dynamics}(b) and Fig.~\ref{fig:dynamics}(c). Thus, inertial effects are found to be important in these non-buoyant wet granular systems, even though the density of the grains are well within 1\% of the density of the fluid. 

Having established the effective friction experienced by the intruder, as a function of speed and the important time scale, we next investigate the observed dynamics from the perspective of the rearrangements of the medium as a result of the intruder motion. 

\section{Intruder driven medium flow}
\begin{figure}
\begin{center}
\includegraphics[width=0.6\textwidth]{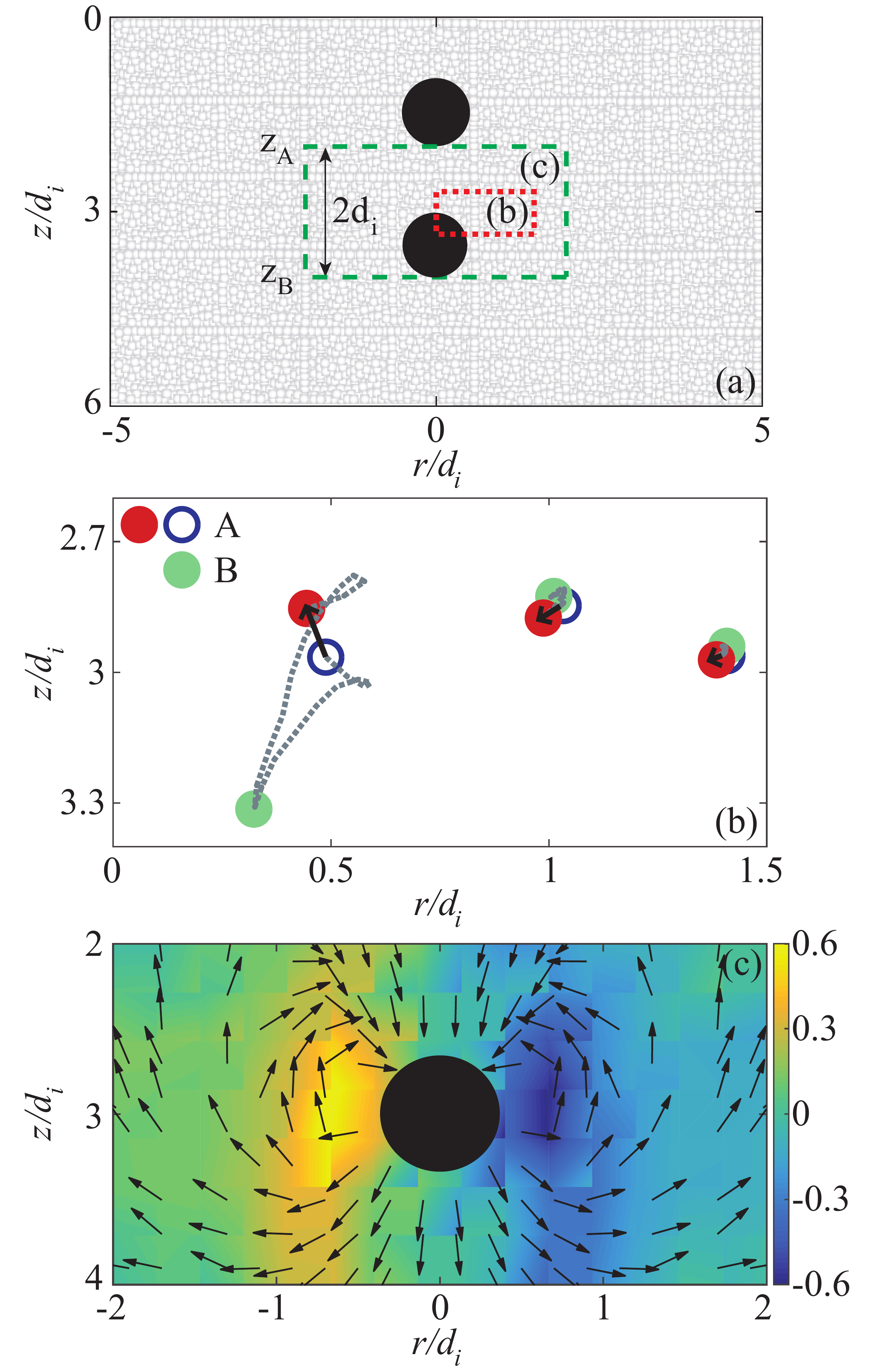}
\end{center}
\caption{(a) Schematic of the experimental system used to measure the medium rearrangements. The boxes represent the displacements shown in (b) and velocity field in (c). (b) Tracers in a vertical plane are observed to follow a systematic trajectory as the intruder is moved from a depth $z_A$ (blue/gray filled circle) to $z_B$ (empty circle) and back up to $z_A$ (green/gray filled circle) as shown in the inset.  The net displacement of the tracers after the cycle is shown by a (gray/red) arrow and is observed to decrease with distance from the intruder.
 (c) The velocity field of the medium and its curl ($v_i = 0.01$\,mm\,s$^{-1}$). The arrows indicate the direction of the flow. The magnitude of the curl is given by the color bar. }
\label{fig:flow}
\end{figure}

For the complementary study of the medium dynamics, we
found it more convenient to modify the experimental system somewhat. We use a container with a rectangular crosssection with height $H_w = 32$\,cm and $H_h = 30$\,cm, and horizontal dimensions 50\,cm and 25\,cm to simplify the visualization shown schematically in Fig.~\ref{fig:flow}(a). Further, we also attach a thin rigid rod to the intruder, and use it to push and pull the intruder with a prescribed speed and through a prescribed depth along the vertical central axis of the container, rather then allowing the intruder to fall in gravity. This protocol enabled us to obtain data under well defined conditions more quickly and more flexibly over a wide range of intruder speeds. Further, it enabled us to examine the reversibility of the flow by measuring the flow when the intruder is moved  back to its original depth. 

In the experiments discussed here, we use $d_i = 5$\,cm and a rod with diameter 5\,mm. Because of the large difference in size, the rod was observed to have negligible impact on the overall trends discussed. We visualize the motion of the medium by adding neutrally buoyant opaque tracer particles with diameter 5\,mm to the medium. This size was choosen to be large enough so we could easily follow the trajectory as the tracer moved with the granular medium, but small enough compared to the gradients in the mean flow. The velocity measurements are performed by moving the intruder vertically from a prescribed depth $z_A$ down to a prescribed depth $z_B$, before returning it back up to its original position as shown schematically in Fig.~\ref{fig:flow}(a).  

Sample trajectories recorded for tracers which are located at increasing horizontal distance $r/d_i$ from the line of motion are shown in Fig.~\ref{fig:flow}(b), corresponding to the red/gray dotted box shown in Fig.~\ref{fig:flow}(a).  Here, the intruder is moved with $v_i = .01$ mm\,s$^{-1}$ from a depth $z_A/d_i =2$ to $z_B/d_i = 4$ before being returned to $z_A$ after a wait time of 20 minutes. This intruder speed corresponds here to the quasi-static limit where the effective friction $\mu_e$ appears constant. Corresponding movies of the motion of the tracers tracked as the intruder is moved down and  back up to its original position at various speeds can be found in the Supplementary Documentation~\cite{supdoc}. The trajectory of the tracers, while not fully periodic is observed to be quite well defined. In this example, close to the intruder, the tracers move away and then get drawn up closer to the center as the intruder moves down. Then, during the second half of the cycle,  the tracers are pushed away and then drawn down as the intruder is returned to its original position. In the representative examples shown, this overall excursion is observed to decrease with distance of the tracers from the line along which the intruder moves.

\subsection{Velocity fields}

\begin{figure}
\begin{center}
\includegraphics[width=0.65\textwidth]{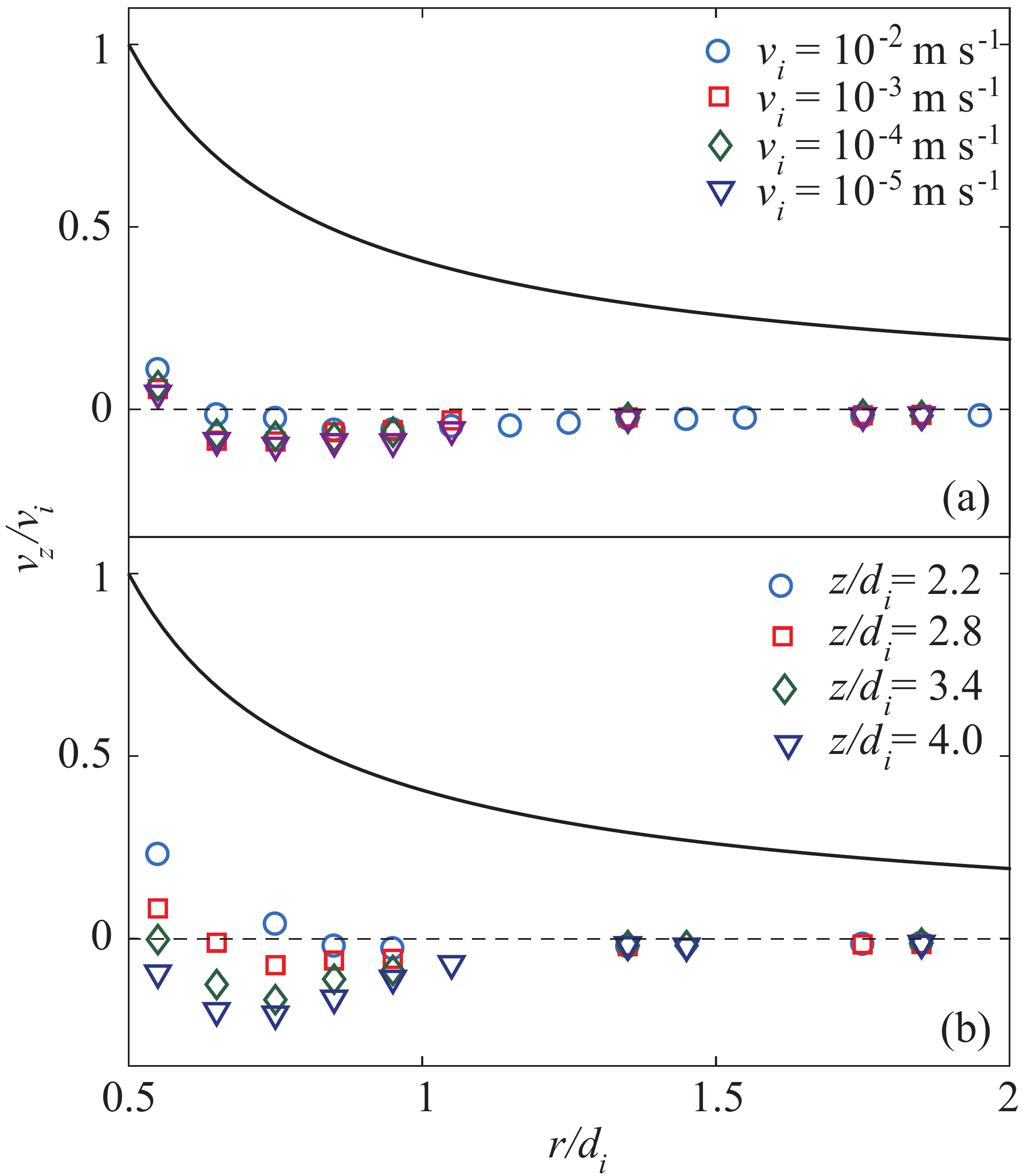}
\end{center}
\caption{(a) The vertical component of the medium velocity $v_z$ as a function of horizontal distance along the equatorial plane of the intruder as it moves down with various speed $v_i$ ($z/d_i = 3$). The velocity of the medium is significantly lower compared to that of a sphere moving with the same speed in a viscous fluid (solid line). (b) $v_z$ as a function of horizontal distance along the equatorial plane of the intruder at various depths ($v_i = 10^{-3}$ m/s). Greater variation is observed with respect to changes in intruder depth compared with intruder velocity. The measurement errors are smaller than the marker size and not drawn for clarity.
}
\label{fig:vel} 
\end{figure}

We obtain the mean flow field around the moving intruder at various speeds using tracer trajectories over a short time interval during which the tracer displacement can be approximated to be linear.  Then, according to our cylindrical coordinate system, the velocity component $v_z$ is along the vertical $z$ axis, and the velocity component $v_r$ along the horizontal distance $r$ from the axis of intruder motion is the same in all the radial directions in the horizontal plane. This is because of the azimuthal symmetry of the flow around the axis of a spherical intruder moving along a line, as well as because the flow decays rapidly compared to the rectangular crosssection of the container.  

A snapshot of the flow field of the medium around the intruder as it descends is shown in Fig.~\ref{fig:flow}(c) using velocity normalized to unity for clarity. Here, the velocity field was obtained by measuring the tracer displacements in a $10^4$ second time interval in which the intruder moves from $z_A$ to $z_B$, and averaging over 10 cycles as the intruder passes through the same depth $z/d_i = 2.5$. To highlight the vorticity of the medium flow, we also superimpose the curl of the velocity field according to the color map which is also shown in Fig.~\ref{fig:flow}(b). One observes from the arrows that the medium is pushed forward along with the intruder directly above and below the intruder, but reverses directions rapidly near the intruder with a vortex-like flow structure near the equatorial plane of the intruder. If one considers the Reynolds Number Re $= \rho_f v_i d_i / \nu$, where $\nu$ is the viscosity of water, then Re $= 0.5$ and laminar flow with no slip at the surface can be expected. If one considers the effect of the hydrogels is to increase the effective viscosity~\cite{shewan15}, then Re would be even lower. Thus, the flow due to the presence of the granular medium appears to be significantly different  compared to that for a viscous Newtonian fluid. Further, the recirculating region and the qualitative flow structure also appear to be different than observed in clay suspensions where a negative wake has been noted~\cite{gueslin06}, and in dry granular medium where cavitation can occur readily behind fast moving intruders~\cite{kolb13}. 

To quantitatively understand the nature of the medium micromechanics, we plot the measured velocity component $v_z$ along the equatorial plane in Fig.~\ref{fig:vel}(a) and Fig.~\ref{fig:vel}(b) as a function over various $v_i$ and  $z$, respectively. For reference, the calculated velocity for an intruder moving through a viscous fluid~\cite{happel}
\begin{equation}
v_z =  v_i\, (\frac{d_i^3}{16x^3} + \frac{3d_i}{4x})
\label{eq:vel}
\end{equation}
is also plotted in Fig.~\ref{fig:vel}(a,b). We observe that the flow of the grains in the medium shows considerable slip near the intruder surface at $r/d_i = 1/2$ in contrast with the viscous fluid case where $v_z = v_i$. The overall form of the medium velocity is similar over a wide range of $v_i$ with a  reverse flow occurring at $r \sim 0.75d_i$. The reversal is observed to occur closer to the intruder and grow stronger as the depth $z$ of the intruder  increases. Thus, we find that the flow in the case of granular medium immersed in a fluid is strongly confined near the moving intruder and considerably different than a Newtonian fluid. 

Further, comparing the observed velocities measured by varying intruder velocity versus intruder depth, one observes that $v_z/v_i$ scales somewhat over 3-orders-of-magnitude in intruder speeds, although the scaled speeds are systematically lower in the case of the slower intruders. But in the case of $v_z/v_i$ measured at various $z$, the data does not collapse with systematic and significant variation with depth. In particular it can be noted that $v_z/v_i$ decreases monotonically at small depths, whereas, at larger z, $v_z/v_i$ decreases rapidly and becomes negative before decaying to zero over the same distance from the intruder center. Thus, a counter flow develops faster and closer to the intruder with increasing overburden pressure. It is noteworthy that the inertial number $I$, in fact varies over three orders of magnitude from  $4.8 \times 10^{-4}$ to $4.8 \times 10^{-1}$, corresponding to the speed variation probed in Fig.~\ref{fig:vel}(a), and $I$ varies less than factor of two from $4.2 \times 10^{-2}$ to $5.6 \times 10^{-2}$, while the depth is varied in Fig.~\ref{fig:vel}(b). Thus, we do not find a collapse of the flow field around the intruder with $I$ as we found in the case of the effective friction in Fig.~\ref{fig:statics}(b).

\subsection{Flow reversibility and plastic deformation}
\begin{figure*}
\begin{center}
\includegraphics[width=0.65\textwidth]{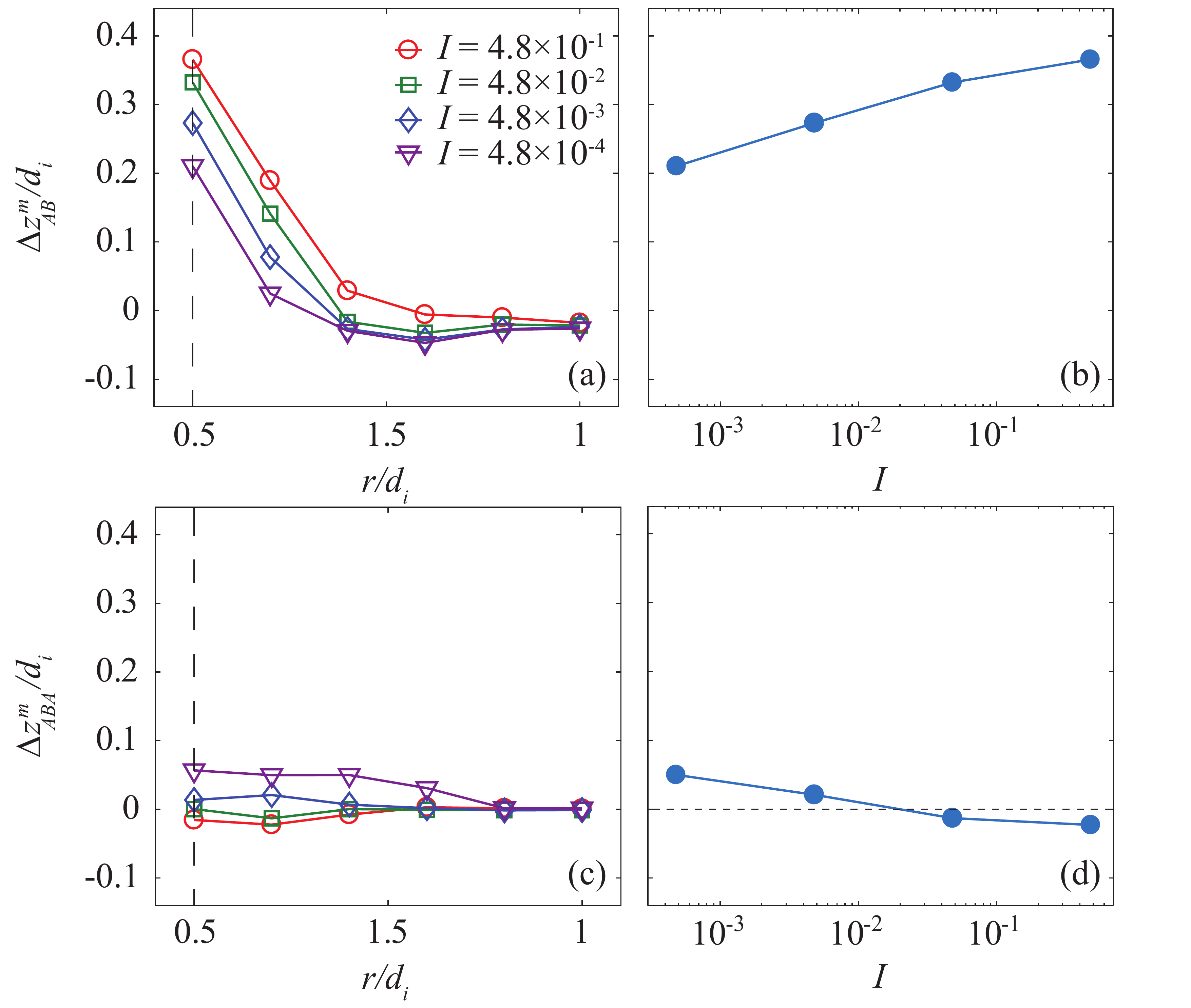}
\end{center}
\caption{(a) Vertical displacement of the medium normalized by the intruder diameter $\Delta z^m_{AB}/d_i$ for various $I$ when the intruder moves down from $z_A$ to $z_B$. (b) The displacements at $r/d_i \approx 0.5$ (indicated by vertical dashed line in (a)) are plotted as a function of $I$ and observed to increase systematically in magnitude.  (c) Vertical displacement of the medium normalized by the intruder diameter $\Delta z^m_{AB}/d_i$ for various $I$ when the intruder moves down from $z_A$ to $z_B$
and then returns back up to original depth $z_A$. (d) The displacement near the intruder corresponding to $r/d_i \approx 0.5$ (the vertical dashed line in (c)) is plotted as a function of $I$ and is observed to decrease and change sign with $I$. The symbols shown in (a) and (c) also correspond to the velocity key in Fig.~\ref{fig:vel}(a). The measurement errors are smaller than the marker size and not drawn for clarity. }
\label{fig:deform}
\end{figure*}

Next, we examine the displacement of the medium as the intruder is moved from $z_A$ to $z_B$, and then also after the intruder is moved back up to its original depth $z_A$ to study the rearrangements of the medium as a result of the fluidization by the intruder.  

The displacement $\Delta z^m_{AB}$ of the tracer particles as the intruder is moved down by $z_B - z_A$ is shown in Fig.~\ref{fig:deform}(a) as a function of distance $r/d_i$ in the horizontal place from the center of the intruder. The data corresponding to varying $v_i$, and thus $I$, are shown in Fig.~\ref{fig:vel} from the quasi-static regime to the inertia dominated regime. One observes the displacement of the medium near the intruder is of order of the radius of the intruder over the range of $v_i$ probed. At lower speeds or lower $I$ one observes that the displacement becomes negative before decaying to zero. But as speed or $I$ is increased, the displacement not only increases overall but stays positive over greater distances. In fact at the highest $I$, $\Delta z^m_{AB}$ appears to simply decay to zero. These trends are consistent with variations of the flow observed by increasing depth $z$ in Fig.~\ref{fig:vel}(b), where increasing depth, which results in lower $I$ also leads to a reversal in the flow. 

To highlight the trend with speed $v_i$, we plot the net displacement $\Delta z^m_{AB}$ of the tracer particles near the intruder $r/d_i \approx 0.5$ in Fig.~\ref{fig:deform}(b) as a function of $I$. One observes that the medium displacement grows systematically with intruder speeds or $I$ as the medium appears to get more fluidized at higher speeds.

Further information is gained by then examining the net displacements of the medium when the intruder is moved back up to its original depth $z_A$. Thus, the intruder is first moved from $z_A$ to \textcolor{black}{$z_B$}, and then after a 20 minute wait, moved back to its original depth $z_A$ with the same speed. We accordingly plot $\Delta z^m_{ABA}$ versus $I$ in Fig.~\ref{fig:deform}(c) and $\Delta z^m_{ABA}$ at $r/d_i \approx 0.5$ in Fig.~\ref{fig:deform}(d). Thus,  examining the displacements over the entire cycle, we find that $\Delta z_{ABA}$ is small overall, while changing from net positive to negative as $I$ increases. The measurements reported here are averaged over 10 different runs and the observed fluctuations are less than the small, but systematic variation, observed here. Thus, while the plastic displacements change systematically from being positive to negative, the overall magnitude remains small even though the inertial number is varied over a wide range by changing the intruder velocity.   
It is noteworthy that $\Delta z_{ABA}$ is not zero even at the lowest speeds, where inertial effects as measured by $I$ are negligible, are consistent with studies of diffusion in sheared suspensions~\cite{pine05}. There it was found that cyclically sheared suspensions with solid particles become irreversible for sufficiently large concentrations due to chaotic particle interactions. {\color{black} Flow reversibility can occur in case of athermal frictionless hard core particles suspended in a fluid in the limit of zero Reynolds number. However, the hysteresis inherent in case of contact between frictional non-buoyant grains can lead to irreversibility even at low speeds because of the sensitivity to initial condition in disordered multi-body systems as we observe here.}   

\section{Conclusions}

In summary, we have developed experiments to measure the friction encountered by an intruder moving through a wet granular medium as a function of its speed and material properties. This system further enables us to visualize the resulting rearrangement of the surrounding medium using direct optical imaging. When the intruder is released at the surface of the medium, it is found to drop slowly and come to rest well above the bottom of the container depending on its size and density. We estimate the drag experienced by the intruder in terms of an effective friction that can be described by a formula with a non-zero yield stress component corresponding to the static limit, and a second component which increases as a power-law with intruder speed corresponding to increasing inertial effects. We find that the system dependence of the friction can be then collapsed onto a single curve using the inertial number rather than the viscous number, even though the density of the grains in the medium is only slightly greater than the fluid. 

By visualizing and measuring the displacement of the medium, significant slip is found near the intruder surface. The flow of the medium is found to be strongly confined close to the intruder in comparison to a viscous fluid, and much smaller in magnitude compared to a viscous fluid. At low speeds, the motion of the medium is found to remain essentially reversible, and then remains so even as the inertial number increases and the effective rheology of the medium changes away from the quasi-static regime. While the effective friction encountered by the intruder depends only on the inertial number, the variation of the medium flow with depth and intruder velocity are not found to be linked via the inertial number, i.e. the velocity profiles corresponding to the same inertial number differ, when observed by varying intruder speed or intruder depth. Nonetheless, it can be observed that medium flow does become increasingly localized, either by decreasing speed, or by increasing depth, as may be anticipated based on their effect on the inertial number. 

Thus, our study provides not only quantitative data on intruder dynamics in sedimented wet granular medium and empirical formulas on the probed rheology, but also perspective on the nature of the resulting unsteady flow of the surrounding medium.

\begin{acknowledgments}
We thank Xavier Clotet for contributing to preliminary experiments, and Benjamin Allen for discussions. Acknowledgment is made of the Donors of
the American Chemical Society Petroleum Research Fund for partial support of this research. This work was also partially supported by the National Science Foundation under Grant No. CBET 1335928, and under Grant No. NSF PHY-1748958.
\end{acknowledgments}

\bibliographystyle{unsrt}

\begin{thebibliography}{10}

\bibitem{gray09}
Murray Gray, Zhenghe Xu, and Jacob Masliyah.
\newblock Physics in the oil sands of alberta.
\newblock {\em Physics Today}, 3:31--35, 2009.

\bibitem{balmforth14}
Neil Balmforth, Ian Frigaard, and Guillaume Ovarlez.
\newblock Yielding to stress: Recent developments in viscoplastic fluid
  mechanics.
\newblock {\em Annual Review of Fluid Mechanics}, 46:121--146, 2014.

\bibitem{pacheco10}
F.~Pacheco-V\'azquez and J.C. Ruiz-Su\'arez.
\newblock Cooperative dynamics in the penetration of a group of intruders in a
  granular medium.
\newblock {\em Nature Communications}, 1:123, 2010.

\bibitem{hosoi15}
A.E. Hosoi and Daniel~I. Goldman.
\newblock Beneath our feet: Strategies for locomotion in granular media.
\newblock {\em Annual Review of Fluid Mechanics}, 47:431--453, 2015.

\bibitem{maladen10}
R.~Maladen, Y.~Ding, P.~Umbanhowar, A.~Kamor, and D.~Goldman.
\newblock Biophysically inspired development of a sand-swimming robot.
\newblock In {\em Proceedings of Robotics: Science and Systems}, Zaragoza,
  Spain, June 2010.

\bibitem{reddy11}
A.~Reddy, Y.~Forterre, and O.~Pouliquen.
\newblock Evidence of mechanically activated processes in slow granular flows.
\newblock {\em Phys. Rev. Lett.}, 106:108301, 2011.

\bibitem{bergmann17}
Philip~J. Bergmann, Kyle~J. Pettinelli, Marian~E. Crockett, and Erika~G.
  Schaper.
\newblock It's just sand between the toes: how particle size and shape
  variation affect running performance and kinematics in a generalist lizard.
\newblock {\em Journal of Experimental Biology}, 220:3706--3716, 2017.

\bibitem{jslonaker17}
James Slonaker, D.~Carrington Motley, Qiong Zhang, Stephen Townsend, Carmine
  Senatore, Karl Iagnemma, and Ken Kamrin.
\newblock General scaling relations for locomotion in granular media.
\newblock {\em Physical Review E}, 95:0529015, 2017.

\bibitem{katsuragi07}
Hiroaki Katsuragi and Douglas~J. Durian.
\newblock Unified force law for granular impact cratering.
\newblock {\em Nature Physics}, 3:420--423, 2007.

\bibitem{katsuragi13}
Hiroaki Katsuragi and Douglas~J. Durian.
\newblock Drag force scaling for penetration into granular media.
\newblock {\em Phys. Rev. E}, 87:052208, 2013.

\bibitem{hilton13}
J.~E. Hilton and A.~Tordesillas.
\newblock Drag force on a spherical intruder in a granular bed at low froude
  number.
\newblock {\em Phys. Rev. E}, 88:062203, 2013.

\bibitem{takehara14}
Yuka Takehara and Ko~Okumura.
\newblock High-velocity drag friction in granular media near the jamming point.
\newblock {\em Phys. Rev. Lett.}, 112:148001, 2014.

\bibitem{takada16}
Satoshi Takada and Hisao Hayakawa.
\newblock Drag law of two-dimensional granular fluids.
\newblock {\em Journal of Engineering Mechanics}, 143:C4016004, 2017.

\bibitem{kumar17}
Sonu Kumar, K.~Anki Reddy, Satoshi Takada, and Hisao Hayakawa.
\newblock Scaling law of the drag force in dense granular media.
\newblock {\em arXiv}, 1712.09057v1, 2017.

\bibitem{happel}
John Happel and Howard Brenner.
\newblock {\em Low Reynolds number hydrodynamics with special applications to
  particulate media}.
\newblock Kluwer, Boston, 1983.

\bibitem{stevens05}
A.B.Stevens and C.M.Hrenya.
\newblock Comparison of soft-sphere models to measurements of collision
  properties during normal impacts.
\newblock {\em Powder Technology}, 154:99--109, 2005.

\bibitem{delannay17}
R.~Delannay, A.~Valance, A.~Mangeney, O.~Roche, and P.~Richard.
\newblock Granular and particle-laden flows: from laboratory experiments to
  field observations.
\newblock {\em J. Phys. D: Appl. Phys.}, 50:053001, 2017.

\bibitem{lauga09}
E.~Lauga and T.~R. Powers.
\newblock The hydrodynamics of swimming microorganisms.
\newblock {\em Reports on Progress in Physics}, 72:096601, 2009.

\bibitem{seto13}
Ryohei Seto, Romain Mari, Jeffrey~F Morris, and Morton~M Denn.
\newblock Discontinuous shear thickening of frictional hard-sphere suspensions.
\newblock {\em Physical Review Letter}, 111:218301, 2013.

\bibitem{brown14}
Eric Brown and Heinrich~M Jaeger.
\newblock Shear thickening in concentrated suspensions: phenomenology,
  mechanisms and relations to jamming.
\newblock {\em Reports on Progress in Physics}, 77:046602, 2014.

\bibitem{apanaitescu17}
Andreea Panaitescu, Xavier Clotet, and Arshad Kudrolli.
\newblock Drag law for an intruder in granular sediments.
\newblock {\em Physical Review E}, 95:03290, 2017.

\bibitem{cruz05}
F.~da~Cruz, S.~Emam, M.~Prochnow, J.-N. Roux, and F.~Chevoir.
\newblock Rheophysics of dense granular materials: Discrete simulation of plane
  shear flows.
\newblock {\em Phys. Rev. E}, 72:021309, 2005.

\bibitem{herschel26}
W.H. Herschel and R.~Bulkley.
\newblock Konsistenzmessungen von gummi-benzollösungen.
\newblock {\em Kolloid Zeitschrift}, 39:291--300, 1926.

\bibitem{hemphill93}
T.~Hemphill, W.~Campos, and A.~Pilehvari.
\newblock Yield-power law model more accurately predicts mud rheology.
\newblock {\em Oil and Gas Journal}, 91:45–50, 1993.

\bibitem{gueslin09}
B.~Gueslin, L.~Talini, and Y.~Peysson.
\newblock Sphere settling in an aging yield stress fluid: link between the
  induced flows and the rheological behavior.
\newblock {\em Rheol Acta}, 48:961--970, 2009.

\bibitem{kann11}
Stefan von Kann, Jacco~H. Snoeijer, Detlef Lohse, and Devaraj van~der Meer.
\newblock Nonmonotonic settling of a sphere in a cornstarch suspension.
\newblock {\em Physical Review E}, 84:060401, 2011.

\bibitem{mukhopadhyay11}
Shomeek Mukhopadhyay and Jorge Peixinho.
\newblock Packings of deformable spheres.
\newblock {\em Phys. Rev. E}, 84:011302, 2011.

\bibitem{apanaitescu14}
Andreea Panaitescu and Arshad Kudrolli.
\newblock Epitaxial growth of ordered and disordered granular sphere packings.
\newblock {\em Physical Review E}, 90:032203, 2014.

\bibitem{silbert02}
Leonardo~E. Silbert, Deniz Ertas, Gary~S. Grest, Thomas~C. Halsey, and Dov
  Levine.
\newblock Geometry of frictionless and frictional sphere packings.
\newblock {\em Phys. Rev. E}, 65:031304, 2002.

\bibitem{supdoc}
See supplemental material at [url will be inserted by publisher] for movie of
  sedimentation.

\bibitem{Brzinski13}
T.~A.~Brzinski III, P.~Mayor, and D.~J. Durian.
\newblock Depth-dependent resistance of granular media to vertical penetration.
\newblock {\em Phys. Rev. Lett.}, 111:168002, 2013.

\bibitem{gmidi04}
GDR MiDi.
\newblock On dense granular flows.
\newblock {\em European Physical Journal E}, 14:341--365, 2004.

\bibitem{boyer11}
F.~Boyer, {\'E}.~Guazzelli, and O.~Pouliquen.
\newblock Unifying suspension and granular rheology.
\newblock {\em Physical Review Letters}, 107, Oct 2011.

\bibitem{shewan15}
Heather~M. Shewan and Jason~R. Stokes.
\newblock Viscosity of soft spherical micro-hydrogel suspensions.
\newblock {\em Journal of Colloid and Interface Science}, 442:75--81, 2015.

\bibitem{gueslin06}
B.~Gueslin, L.~Talini, B.~Herzhaft, Y.~Peysson, and C.~Allain.
\newblock Flow induced by a sphere settling in an aging yield-stress fluid.
\newblock {\em Phys. Fluids}, 18:103101, 2006.

\bibitem{kolb13}
Evelyne Kolb, Pierre Cixous, Niels Gaudouen, and Thierry Darnige.
\newblock Rigid intruder inside a two-dimensional dense granular flow: Drag
  force and cavity formation.
\newblock {\em Phys. Rev. E}, 87:032207, 2011.

\bibitem{pine05}
D.~J. Pine, J.~P. Gollub, J.~F. Brady, and A.~M. Leshansky.
\newblock Chaos and threshold for irreversibility in sheared suspensions.
\newblock {\em Nature}, 438:997--1000, 2005.

\end{thebibliography}

\end{document}